\documentstyle[12pt,preprint,aps,epsfig]{revtex}
\renewcommand{\baselinestretch}{1.25}
\begin{document}
\draft
\title{
Ground-state phases  
in a system of two competing square-lattice Heisenberg antiferromagnets
}
\author{D.Schmalfu{\ss}, R.Herms, J.Richter and J.Schulenburg
}
\address{
Institut f\"ur Theoretische Physik, Otto-von-Guericke Universit\"at,
Magdeburg,\\
P.O.B. 4120, 39016 Magdeburg, Germany\\
}

\maketitle

\begin{abstract}
We study a two-dimensional (2D) spin-half Heisenberg model
related to the  quasi 2D antiferromagnets $(Ba,Sr)_2Cu_3O_4Cl_2$ by means of
exact diagonalization and spin-wave theory.
The model consists of two inequivalent interpenetrating square-lattice
Heisenberg antiferromagnets $A$ and $B$. 
While the antiferromagnetic interaction $J_{AA}$ within the $A$ subsystem 
is  strong the coupling $J_{BB}$ within the $B$ subsystem is much weaker.
The coupling $J_{AB}$ between A and B subsystems is competing 
giving rise for interesting frustration effects. 
In dependence of the strength of $J_{AB}$ we find a collinear 
N\'{e}el phase, non-collinear states with zero magnetizations as well as
canted and collinear ferrimagnetic phases with non-zero magnetizations.  
For not too large 
values of frustration $J_{AB}$, which correpond to the situation in 
$(Ba,Sr)_2Cu_3O_4Cl_2$,
we  have N\'{e}el
 ordering in both subsystems A and B. In the classical limit
these two N\'{e}el
 states are decoupled.
Quantum fluctuations lead to 
a fluctuational coupling between both subsystems ('order from
disorder') and select the collinear structure. 
For stronger  $J_{AB}$ we find evidence for a novel spin state with
coexisting N\'{e}el ordering in the $A$ subsystem and disorder in the $B$
subsystem. 
\end{abstract}

\section{Introduction} \label{I}
\begin{figure}
\begin{center}
\epsfig{file=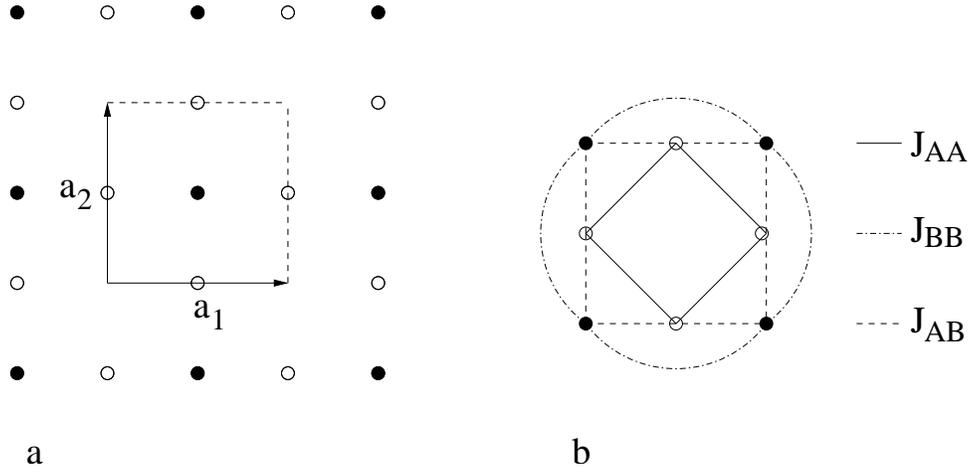,scale=0.6,angle=0.0}
\end{center}
\caption{a: The A-B lattice with its geometrical unit cell defined by the 
unit vectors
$ {\bf {a}}_{1}=\left(a,0\right)$ and $ {\bf
{a}}_{2}=\left(0,a\right)$. 
Empty(filled) circles denote A(B)-spins, respectively.
\newline
b: Exchange couplings of the model 
(\ref{eq1}). 
\newline
\label{fig1}}
\end{figure}

The exciting collective magnetic properties of layered cuprates have
attracted much attention over the last decade. 
A lot of activity in this field
was stimulated by the possible connection of spin fluctuations with the
phenomenon of high-temperature superconductivity. But, the rather unusual
properties of quantum magnets deserve study on their own to gain a deeper
understanding of these quantum many-body systems. In recent years some of those materials
such as $Ba_{2}Cu_{3}O_{4}Cl_{2}$ and $Sr_{2}Cu_{3}O_{4}Cl_{2}$ 
have been studied experimentally and theoretically in more detail
\cite{ref4,ref5,ref3,ref1,ref2}. The most important difference between
 $(Ba,Sr)_2Cu_3O_4Cl_2$  and their
parent compound $La_{2}CuO_{4}$ is the existence of additional
$Cu(B)$-atoms located at the centre of every second
$Cu(A)$-plaquette. Both subsystems $A$ and $B$ form square lattices, 
however,
with different orientations and lattice constants.
This A-B lattice is illustrated in  Fig.\ref{fig1}a. Both
copper sites $Cu(A)$ and $Cu(B)$ carry spin half, i.e. quantum fluctuations
are important.
Since the magnetic couplings $J_{AA}$ between A spins as well as $J_{BB}$
between B spins are antiferromagnetic and the coupling $J_{AB}$
between A and B spins is
frustrating we have a system of two competing
antiferromagnetic spin-half subsystems. 

Three-dimensional  examples of two interpenetrating antiferromagnets 
like garnets 
$Mn_3Cr_2Ge_3O_{12}$ or
 $(Fe_xGa_{1-x})_2Ca_3Ge_3O_{12}$ 
were discussed by several authors (see e.g. \cite{valya,shender,brueck}). 
For the quasi two-dimensional cuprates like 
$(Ba,Sr)_2Cu_3O_4Cl_2$
the quantum fluctuations 
are more important than in the three-dimensional garnets and 
the interplay of competing interactions 
with strong quantum fluctuations may lead to interesting
magnetic phenomena. 

Noro {\em et al.} \cite{ref4} reported 
two magnetic phase transitions at $T_A = 320 K$ and
at $T_B=40 K$ for $Ba_2Cu_3O_4Cl_2$
being attributed to respective
antiferromagnetic ordering of the $Cu(A)$ and $Cu(B)$ spins. Both critical
temperatures differ in one order of magnitude indicating a strongly
antiferromagnetic coupling between $Cu(A)$ spins and a comparatively small
antiferromagnetic coupling between $Cu(B)$ spins, which is confirmed by
band-structure calculations \cite{rosner}. 
According to
Chou {\em et al.} \cite{ref3} the weak ferromagnetic moment found
experimentally \cite{ref5} could be
understood as a consequence of bond-dependent interactions such as 
pseudodipolar
couplings. 

The minimal model to describe the main magnetic properties of the competing
antiferromagnets on the A-B lattice 
is the antiferromagnetic Heisenberg model with three exchange 
couplings   $J_{AA}$, $J_{BB}$
and $J_{AB}$. In what follows we call this model A-B model.
Some preliminary results for a finite system of ${\cal N}=24$ sites
were reported in the conference paper \cite{ref10}.  
However, to describe the weak ferromagnetism observed in these compounds 
anisotropic interactions seem to be needed \cite{ref3,ref1,ref2}.

In this paper we want to study the influence of strong quantum
fluctuations and frustration on the ground state of the A-B model using
spin-wave theory and exact diagonalization.
The paper is organized as follows: In Section \ref{II} 
we introduce the A-B model 
and illustrate the classical magnetic ground-state phases 
in the considered parameter region. In Section \ref{III} we present an 
exact-diagonalization study of the ground-state phases
and in Section \ref{IV} the linear 
spin-wave approach is used to analyse the N\'{e}el phase realized for small
$J_{AB}$ in more detail.
In Section \ref{V} a summary is
given.

\section{The A-B model and its classical ground-state phases}
\label{II}
We consider the Hamiltonian (cf. Fig. \ref{fig1}b)
\begin{equation}
H=J_{AA}\sum_{\left\langle m\in A,\, n\in A\right\rangle}{\bf {S}}_{m}
\cdot{\bf
{S}}_{n}+J_{BB}\sum_{\left\langle m\in B,\, n\in B\right\rangle}{\bf
{S}}_{m}\cdot
{\bf
{S}}_{n}+J_{AB}\sum_{\left\langle m\in A,\, n\in B\right\rangle}{\bf {S}}_{m}
\cdot{\bf
{S}}_{n},
\label{eq1}
\end{equation}
where the sums run over neighbouring sites only. 
$J_{AA}$ and $J_{BB}$ denote the
antiferromagnetic couplings within the $A(B)$-subsystems, respectively. 
We focus our discussion  on parameters $J_{AA}=1$ and $J_{BB}=0.1$ 
which corresponds to the
situation in $(Ba,Sr)_2Cu_3O_4Cl_2$.
The value of the frustrating inter-subsystem coupling $J_{AB}$ is less
reliably known. We 
consider antiferromagnetic $J_{AB}$ and use it as the free
parameter of the model. 
The lattice
consists of ${\cal N}= 3N$ spins with three spins per geometrical unit cell 
and ten  couplings in it.

\begin{figure}
\begin{center}
\epsfig{file=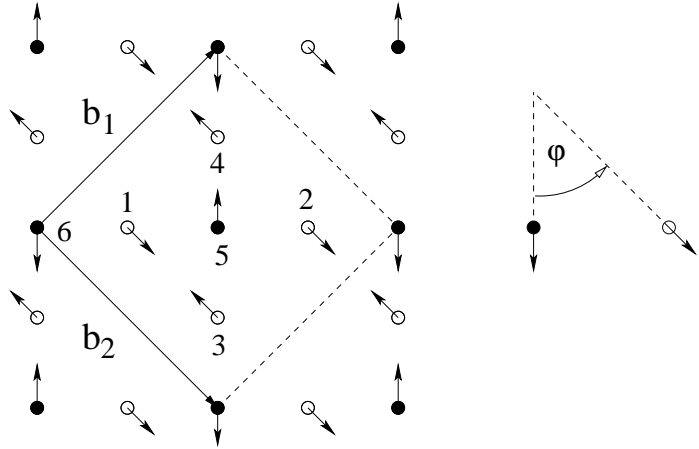,scale=0.6,angle=0.0}
\end{center}
\caption{
The classical ground state for
$J_{AB}< 2\sqrt{J_{AA}J_{BB}}$ (N\'{e}el phase I, cf. table \ref{tab1}).
Both subsystems 
possess N\'eel order. Because of the vanishing classical mean field both subsystems 
decouple magnetically and can freely rotate with respect to 
each other. 
The angle $\varphi$ is parametrizing this degree of freedom. 
The magnetic unit cell containing
six spins is given by $ {\bf {b}}_{1\left(2\right)}= {\bf {a}}_{1}\pm{\bf
{a}}_{2}$. The spins within the unit cell are labeled by a running index
$n=1,\ldots,6$ corresponding to the six different magnons to be 
introduced in spin-wave theory. 
\label{fig1b}}
\end{figure}

We start with the discussion of the classical ground state, i.e. the spins 
${\bf {S}}_{n}$ are considered as classical vectors of length $s$.
Varying
$J_{AB}$ we have altogether five ground-state phases, see table \ref{tab1}.
Two of them (I and III) have planar spin arrangement, two (II and IV) 
are non-planar and one (V) is collinear. Without loss of
generality we choose in this section 
for the description of planar spin ordering 
the x-y plane.
We start from weak inter-subsystem coupling $J_{AB} \stackrel{>}{\sim} 0$. 
Then we have N\'eel ordering in both
subsystems (phase I, table \ref{tab1}).
These two classical N\'eel states shown in Fig.\ref{fig1b} are 
decoupled and 
can rotate freely with
respect to each other, i.e. the ground state is highly degenerated
and this degree of freedom is parametrized by the angle $\varphi$.  
The corresponding magnetic unit cell contains six spins.  
(Thus, later in Section \ref{IV} we have to
introduce six different magnons in the spin-wave theory for this phase.)

\begin{figure}
\begin{center}
\epsfig{file=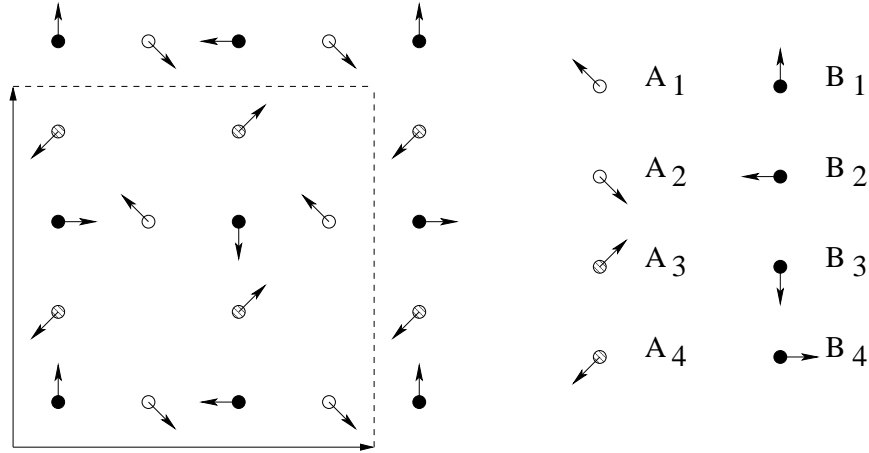,scale=0.6,angle=0.0}
\end{center}
\caption{The spin orientations of phases II, III and IV.
In the non-planar phase II there is an out-of-plane z component of the A 
spins illustrated
by open and dashed circles, 
where the open circles and the dashed circles belong to opposite 
directions of the z
component. In the planar phase III this out-of-plane z component is zero.
In the non-planar phase IV the
out-of-plane z component of A spins as well as of the B spins 
is uniform but opposite to each other.
The magnetic unit cell of the magnetic states of phases II,III,IV 
contains twelve spins.
\label{fig2}}
\end{figure}

At
$J_{AB}=2\sqrt{J_{AA}J_{BB}}$ there is a first-order 
transition  from the N\'eel phase I
to the non-planar ground-state phase II. 
Phase II is illustrated in
Fig.\ref{fig2}. The corresponding magnetic unit cell contains 12 pins 
and is therefore
twice as large as the magnetic unit cell of the N\'eel phase I. In this
state we have eight different spin orientations characterized as follows
\begin{eqnarray}
{\bf {S}}^{II}_{A_{1}}&=&s\left(-\frac {\sqrt{2}}{2}\cos\left(\alpha\right),-\frac
{\sqrt{2}}{2}\cos\left(\alpha\right),\sin\left(\pm\alpha\right)\right),\nonumber\\
{\bf {S}}^{II}_{A_{2}}&=&s\left(\frac {\sqrt{2}}{2}\cos\left(\alpha\right),\frac
{\sqrt{2}}{2}\cos\left(\alpha\right),\sin\left(\pm\alpha\right)\right),\nonumber\\
{\bf {S}}^{II}_{A_{3}}&=&s\left(-\frac {\sqrt{2}}{2}\cos\left(\alpha\right),\frac
{\sqrt{2}}{2}\cos\left(\alpha\right),-\sin\left(\pm\alpha\right)\right),\nonumber\\
{\bf {S}}^{II}_{A_{4}}&=&s\left(\frac {\sqrt{2}}{2}\cos\left(\alpha\right),-\frac
{\sqrt{2}}{2}\cos\left(\alpha\right),-\sin\left(\pm\alpha\right)\right),\nonumber\\
{\bf {S}}^{II}_{B_{1}}&=&s\left(-1,0,0\right),\quad
{\bf {S}}^{II}_{B_{2}}=s\left(0,-1,0\right),\nonumber\\ 
{\bf {S}}^{II}_{B_{3}}&=&s\left(1,0,0\right),\quad
{\bf {S}}^{II}_{B_{4}}=s\left(0,1,0\right),
\label{eq3}
\end{eqnarray}
where $\alpha$ is given by $\alpha=\mbox{arcos}\left(J_{AB}/\sqrt{8}J_{AA}\right)$.
Obviously neighbouring B spins are perpendicular to each other and
consequently the energy $E_{II}$ is independent of $J_{BB}$ (see 
table \ref{tab1}). The in-plane xy components of the A spins 
of neigboring spins are also perpendicular, however, there are 
finite off-plane z
components. These off-plane components (proportional to $\sin
\alpha$, see eq. (\ref{eq3})) decrease with $J_{AB}$ and  
become zero at $J_{AB}=2\sqrt{2} \; J_{AA}$, i.e. 
we have a second-order
transition from the non-planar phase II to the planar phase III at this
point.
The spin
orientations of phase III are given by eq. (\ref{eq3}), too, but with the
additional condition $\alpha=0$ (see Fig. \ref{fig2}).  
In phase III neighbouring  $A$-spins as well as neighbouring $B$-spins are 
perpendicular to each other and consequently the energy depends on $J_{AB}$
only. All three phases I,II,III have a zero net magnetization $S_{total}=0$.

Further increasing $J_{AB}$ favours an antiparallel
 alignment of the A spins relative to the B spins and 
the planar phase III gives way to a  non-planar phase IV where the in-plane 
xy components are aligned as in phases II and III (see Fig. \ref{fig2}). The 
off-plane z components in phase IV  are given as 
$S^{z}_{n \in A}=
+ s \cos \theta$ for A spins and as 
$S^{z}_{n \in B}=-s \; 
\sqrt{1-A^2\sin^2\theta}$ for B spins.
This phase IV can be denoted as canted ferrimagnet \cite{ferri}
and
has a net magnetic moment $S_A$ in the A subsystem 
and $S_B$ in the B subsystem 
resulting in a  finite total magnetic moment $S_{total}$
\begin{eqnarray} \label{IV_2}
\frac{S^{IV}_A}{s2N}=\sqrt{1-\sin^2{\theta}}
\; ; \; 
\frac{S^{IV}_B}{sN}=\sqrt{1-A^2\sin^2{\theta}}
\; ; \; 
{S^{IV}_{total}}= \left |S_A - S_B \right |
\end{eqnarray}
with
\begin{eqnarray}\label{IV_5}
{\sin{\theta}}=
\sqrt{
\frac
{1-\left(\frac{J_{AA}}{J_{AB}}+\frac{J_{AB}}{4J_{BB}}
-\frac{1}{4}\sqrt{-24\frac{J_{AA}}{J_{BB}}+16\frac{J_{AA}^2}{J_{AB}^2}+\frac{J_{AB}^2}{J_{BB}^2}}
\right)^2}
{-3\frac{J_{AA}}{J_{BB}}+2\frac{J_{AA}^2}{J_{AB}^2}+\frac{J_{AB}^2}{8J_{BB}^2}
+\left(\frac{3J_{AA}}{2J_{AB}}-\frac{J_{AB}}{8J_{BB}}\right)
\sqrt{
-24\frac{J_{AA}}{J_{BB}}+16\frac{J_{AA}^2}{J_{AB}^2}+\frac{J_{AB}^2}{J_{BB}^2}} }
}
\end{eqnarray}
and
\begin{eqnarray} \label{IV_6}
A=\frac{J_{AB}}{2\sqrt{2}J_{BB}}-\sqrt{2}\frac{J_{AA}}{J_{AB}}
-\sqrt{-3\frac{J_{AA}}{J_{BB}}+2\frac{J_{AA}^2}{J_{AB}^2}+
\frac{J_{AB}^2}{8J_{BB}^2}}.
\end{eqnarray}
This total moment increases with $J_{AB}$.  
The energy of phase IV is given by 
\begin{eqnarray} \label{IV_1} 
\frac{E_{IV}}{s^2 {\cal N}}&=&
  \frac{4}{3}J_{AA}\left(1-\sin^2\theta_0\right)
+\frac{2}{3}J_{BB}\left(1-A^2\sin^2\theta_0\right)\nonumber \\
&&
-\frac{4}{3}J_{AB}\left(\frac{A}{\sqrt{2}}+
\sqrt{ \left(1-\sin^2\theta_0\right)\left(1-A^2\sin^2\theta_0\right)}\right)
\end{eqnarray}
The phase boundary of the second order phase transition between III and IV
is given by
\begin{equation} \label{IV_7}
 J^{III-IV}_{AB}=\sqrt{2}J_{AA}+\frac{J_{BB}}{\sqrt{2}}
+\frac{1}{2}\sqrt{8J^2_{AA}+24J_{AA}J_{BB}+2J^2_{BB}}
\end{equation}
and yields $J^{III-IV}_{AB}=3.099$ for $J_{AA}=1$ and $J_{BB}=0.1$.

Finally, for large $J_{AB}$ 
the A and B spins are fully polarized along the z axis, i.e.
${\bf S}_{n \in A}=(0,0,+ s)$ and ${\bf S}_{n \in B}=(0,0,-s)$ and 
a collinear  ferrimagnetic phase V
is realized.
The phase boundary of this second order phase transition between IV and V
is given by 
\begin{equation} \label{IV_8}
J^{IV-V}_{AB}=2J_{AA}+J_{BB}+\sqrt{4J^2_{AA}+J^2_{BB}}
\end{equation}
leading to $J^{IV-V}_{AB}=4.102$ for $J_{AA}=1$ and $J_{BB}=0.1$.

\section{The quantum ground state - exact diagonalization}
\label{III}
\begin{figure}
\begin{center}
\epsfig{file=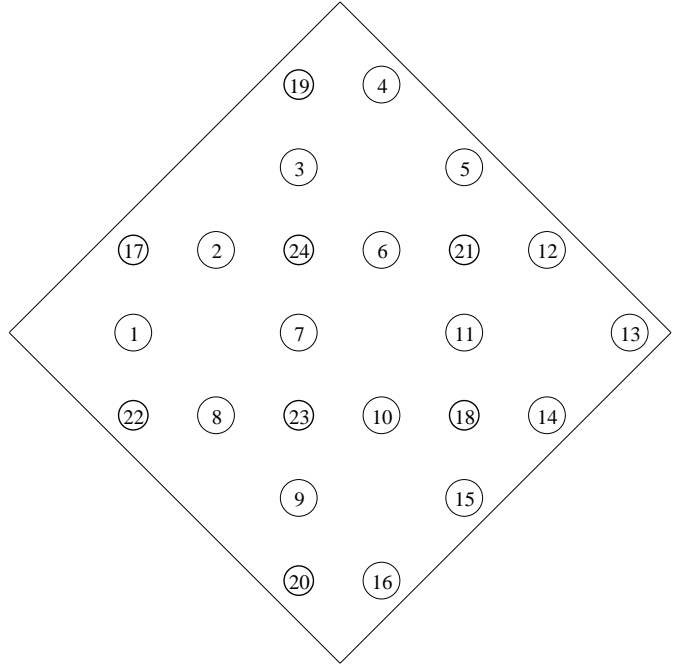,scale=0.55,angle=0.0}
\end{center}
\caption{The finite A-B lattice with ${\cal N}=3N=24$ spins (periodic boundary
conditions). The large circles (sites 1, ..., 16) belong to subsystem A and
the small circles to (sites 17, ..., 24) to subsystem B.\label{fig3}}
\end{figure}

To discuss the influence of quantum fluctuations on the classical phases
studied in the last section  
we use the Lanczos algorithm to calculate the quantum ground state 
of the Hamiltonian (\ref{eq1}) 
for a finite lattice of ${\cal N}= 24$ spins (Fig.\ref{fig3}). Again 
we choose the parameters $J_{AA}=1$, $J_{BB}=0.1$ appropriate for 
$(Ba,Sr)_2Cu_3O_4Cl_2$ and consider $J_{AB}$ as the free parameter.  
For ${\cal N}=24$ sites we have $16$ $A$ spins and $8$ $B$ spins. 
Since the maximal  
magnetic unit cell of the classical ground states contains $12$ spins 
the ${\cal N}=24$ system has the full symmetry of
the classical ground state in the considered parameter region. 
To reduce the
Hilbert space we used all possible translational and point symmetries of the
A-B lattice as
well as spin inversion. The use of the symmetry allows to classify the
different quantum ground states by their symmetry.  

\begin{figure}
\begin{center}
\epsfig{file=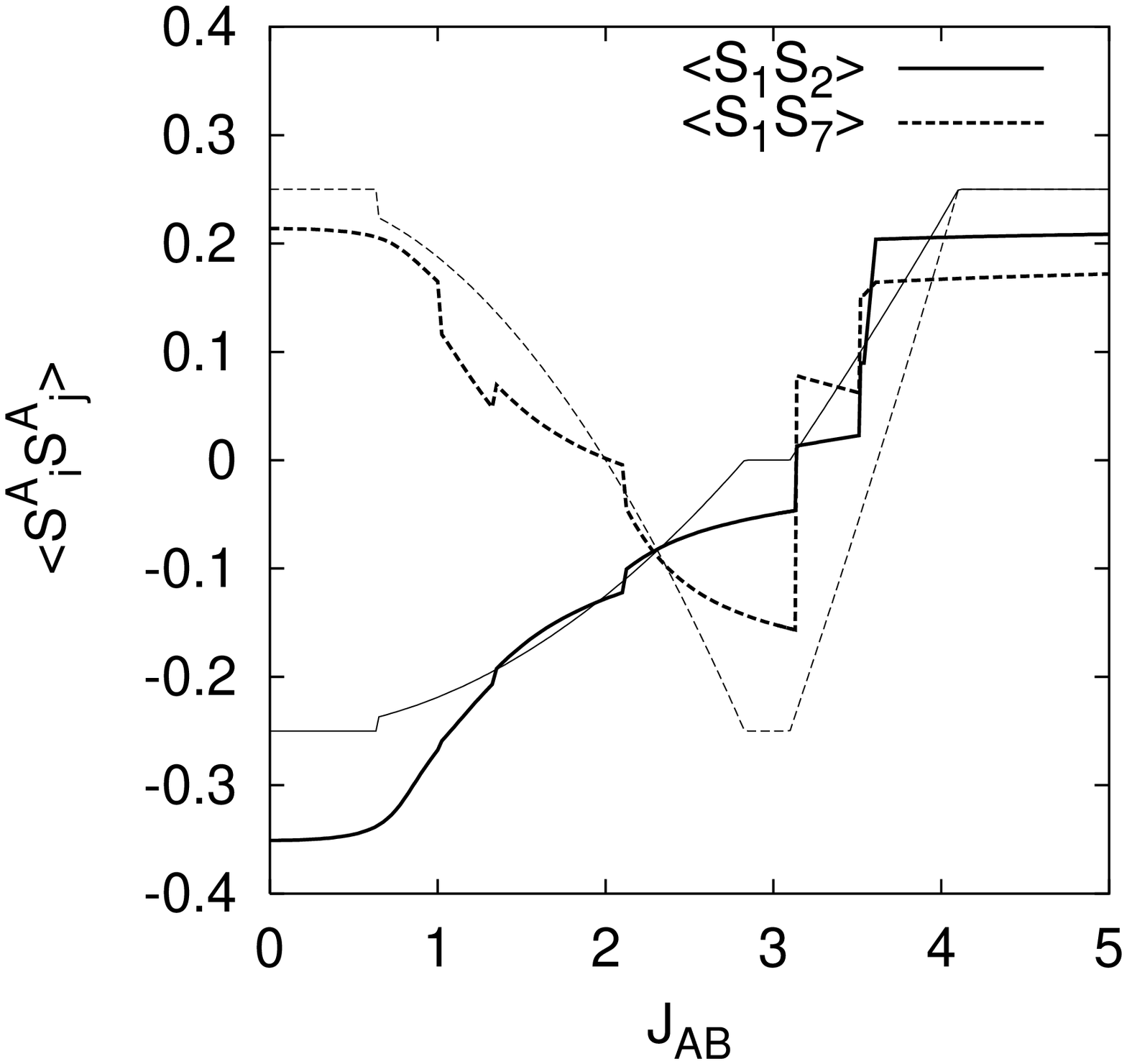,scale=0.6,angle=270.0}
\end{center}
\caption{A-A spin correlation $\langle  {\bf S}^A_i{\bf S}^A_j \rangle$ 
for $(i,j)=(1,2)$ and $(i,j)=(1,7)$ (see Fig. \ref{fig3})
for the classical (thin lines, length of classical
spin vectors is choosen as $s=1/2$) and the quantum (thick lines)
model (${\cal N}=24$, $J_{AA}=1$, $J_{BB}=0.1$). 
\label{figed2}}
\end{figure}

\begin{figure}
\begin{center}
\epsfig{file=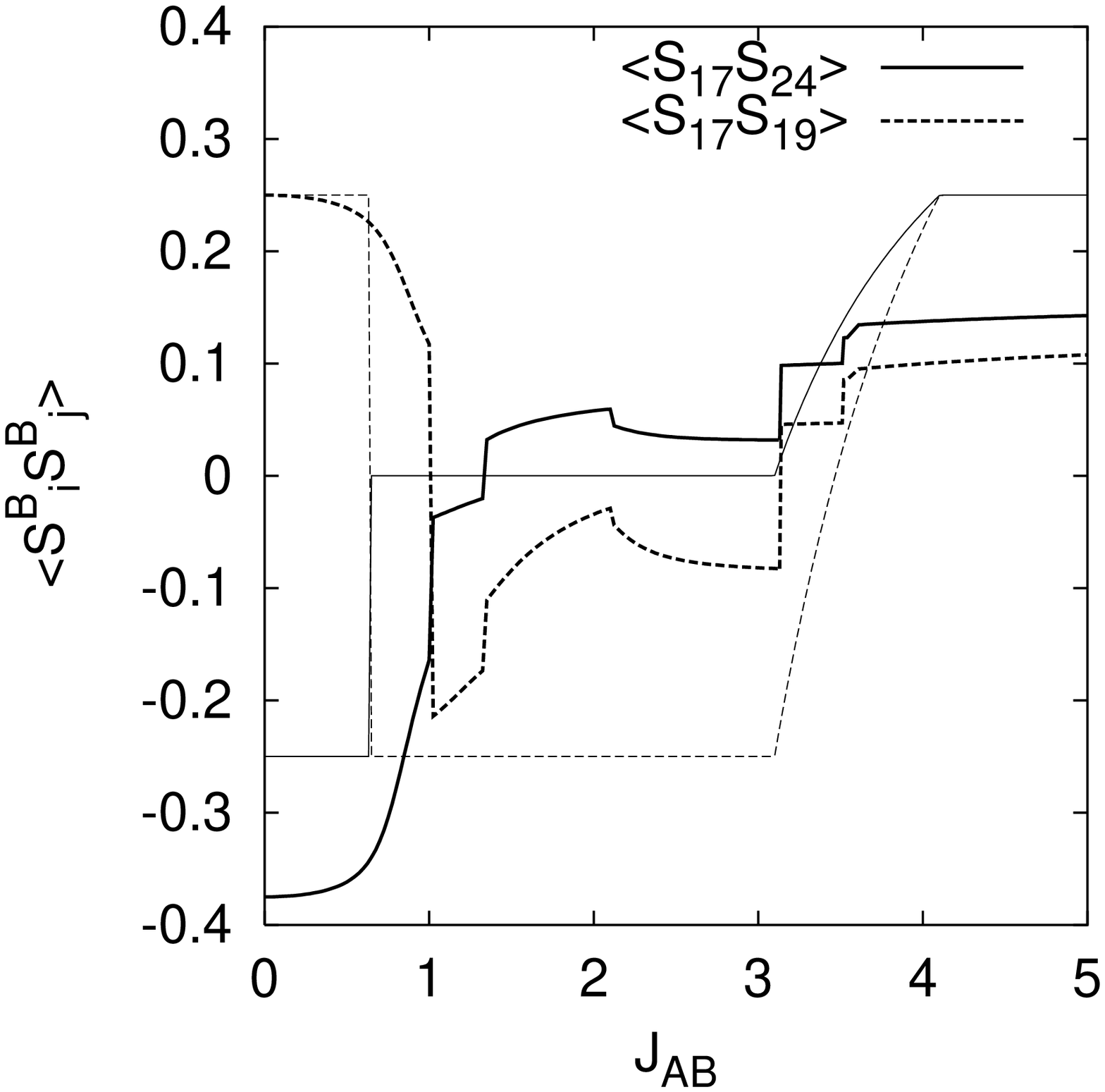,scale=0.6,angle=270.0}
\end{center}
\caption{B-B spin correlation $\langle  {\bf S}^B_i{\bf S}^B_j \rangle$ 
for $(i,j)=(17,24)$ and $(i,j)=(17,19)$ (see Fig. \ref{fig3})
for the classical (thin lines, length of classical
spin vectors is choosen as $s=1/2$) and the quantum (thick lines)
model (${\cal N}=24$, $J_{AA}=1$, $J_{BB}=0.1$).\label{figed3}}
\end{figure}

\begin{figure}
\begin{center}
\epsfig{file=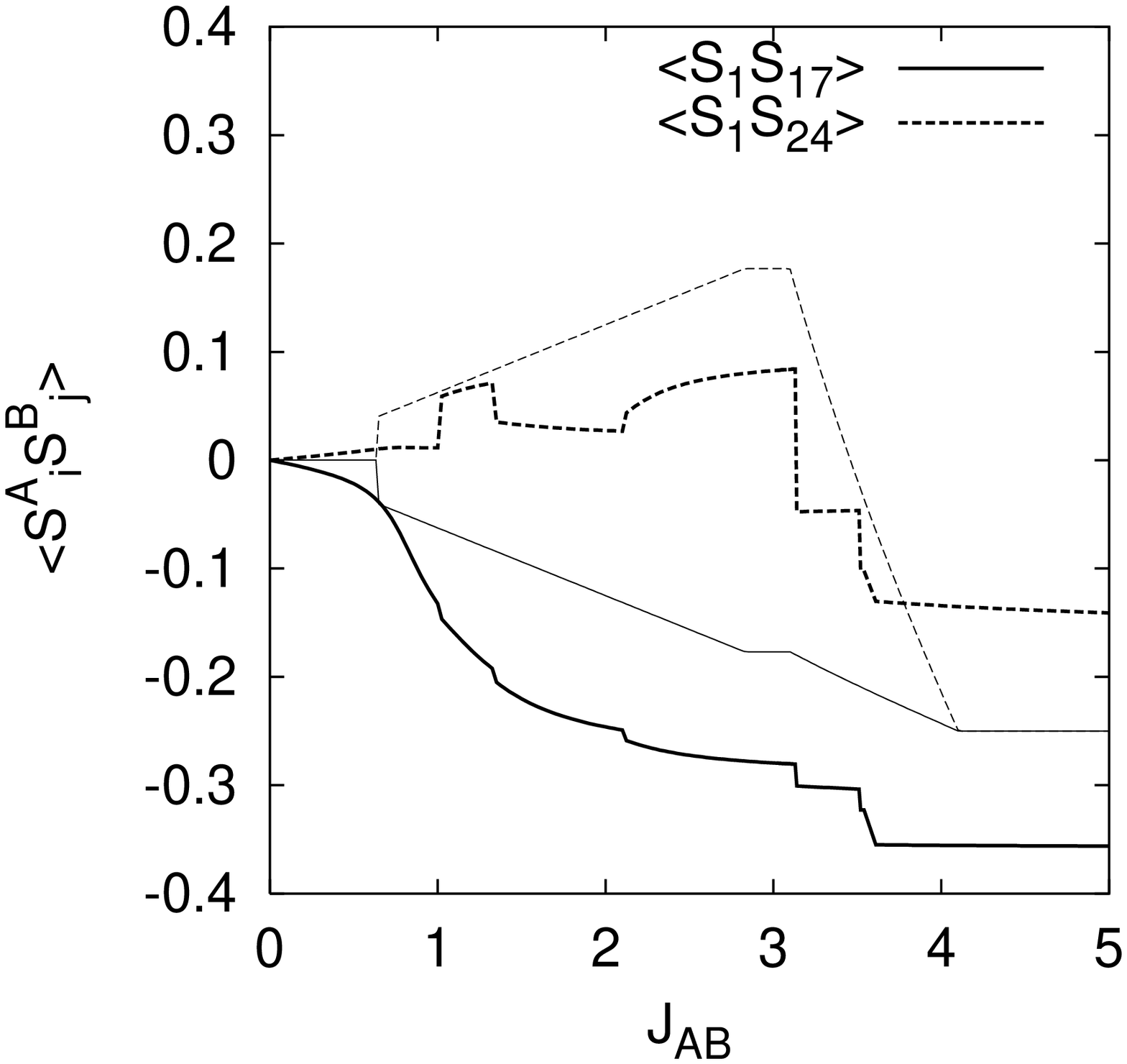,scale=0.6,angle=270.0}
\end{center}
\caption{A-B spin correlation $\langle  {\bf S}^A_i{\bf S}^B_j \rangle$ 
for $(i,j)=(1,17)$ and $(i,j)=(1,24)$ (see Fig. \ref{fig3})
for the classical (thin lines, length of classical
spin vectors is choosen as $s=1/2$) and the quantum (thick lines)
model (${\cal N}=24$, $J_{AA}=1$, $J_{BB}=0.1$).\label{figed4}}
\end{figure}

To compare classical and quantum ground-state phases we present 
the spin-spin correlations in Figs. \ref{figed2}, \ref{figed3}
and \ref{figed4}. The N\'{e}el phase I is present also in the
quantum model. However, the quantum fluctuations lift the classical
degeneracy and both subsystems couple. The fluctuational coupling is known
as {\it order from disorder} effect \cite{villain,shender} and selects a
collinear quantum state with a
finite  A-B spin correlation in the quantum N\'{e}el
phase I (see Fig. \ref{figed4}).    
Moreover, the transition to phase II is shifted to higher 
values of $J_{AB}$ indicating that quantum fluctuations favour 
collinear versus
non-collinear states (see e.g. \cite{ccm}).
The frustrating coupling weakens the A-A and B-B spin correlations in the 
N\'{e}el phase I;
this weakening is stronger for the B-B correlations than for the A-A
correlations. Hence, a disordered quantum ground state similar to the
$J_1-J_2$ model \cite{squa_ref0,squa_ref1,squa_ref2,squa_ref3,squa_ref4}
seems to be possible and will be discussed in more detail in next
Section. 

\begin{figure}
\begin{center}
\epsfig{file=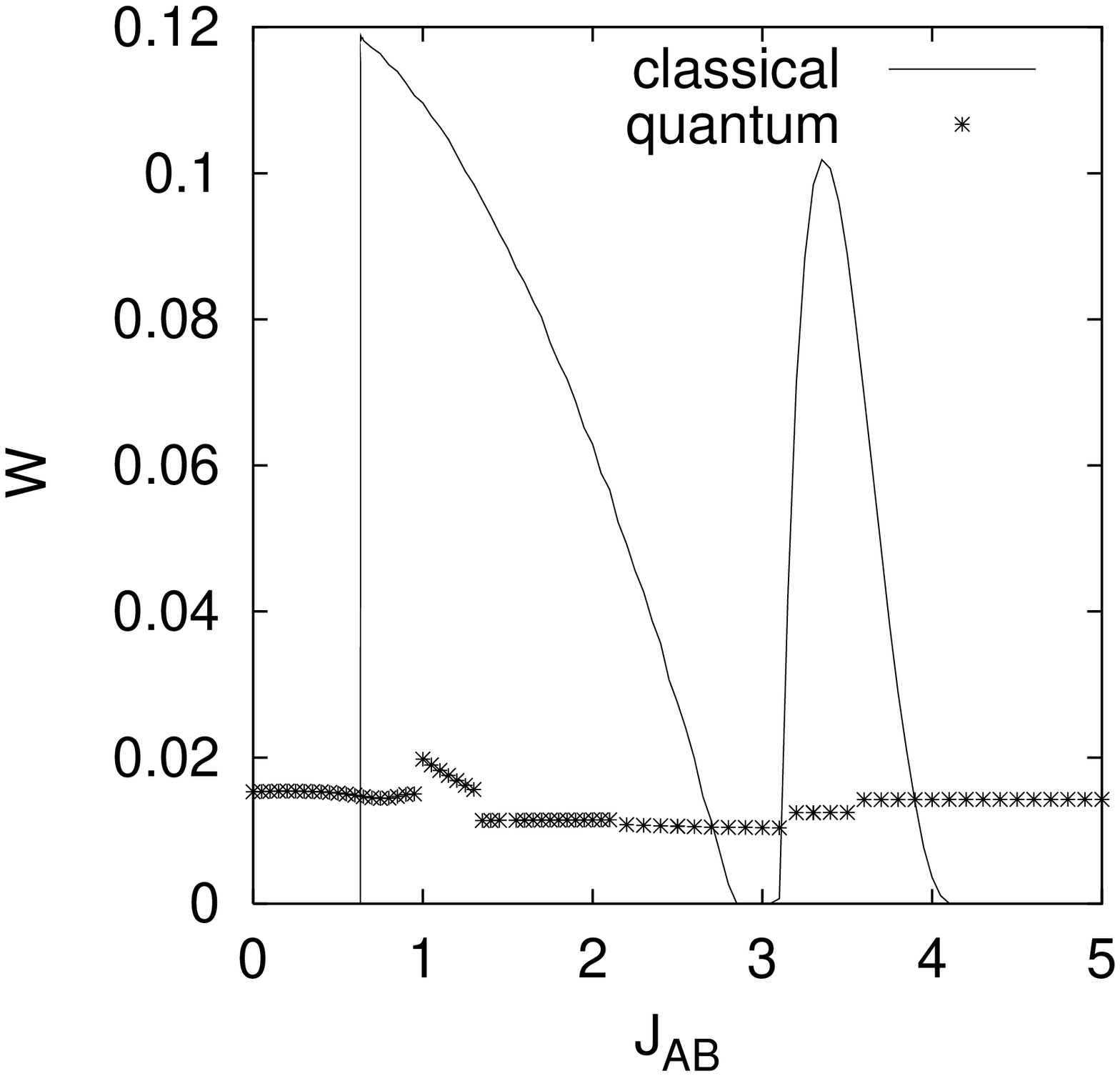,scale=0.6,angle=270.0}
\end{center}
\caption{Scalar chirality $W$ 
(see eq. (\ref{chir}))
for the classical (solid line, length of classical
spin vectors is choosen as $s=1/2$) and the quantum (crosses)
model (${\cal N}=24$, $J_{AA}=1$, $J_{BB}=0.1$).\label{figed5}}
\end{figure}

The quantum N\'{e}el phase  gives way to a fairly
complex spin state at $J_{AB} \approx 1.02$ up to  $J_{AB} \approx 3.13$. 
This state is also a singlet $S=0$ 
as the quantum  N\'{e}el state. The A-A and the A-B spin correlations
of the quantum model follow qualitatively the classical curves (see Figs.
\ref{figed2} and \ref{figed4}). However, we see
some jumps in the correlations connected with level crossings of ground
states belonging to different lattice symmetries. Most likely these level
crossings may be attributed to finite-size effects.  
The change of A-A and A-B 
correlations at $J_{AB} \approx 1.02$ is small. However, the
B-B correlations change strongly at this point. Contrary to the
classical model where the nearest-neighbour B-B correlation is zero and the
next-nearest-neighbour B-B correlation is strongly antiferromagnetic. The
corresponding correlations in the quantum model are both different from zero
and are of the same order of magnitude. One could argue that quantum
fluctuations favour planar versus non-planar arrangement of spins. This
argument is supported by (i) the circumstance that there is a planar classical
state of almost the same energy as the non-planar state having 
finite nearest-neighbour B-B correlations and 
(ii) by investigation of the so-called scalar
chirality $W_{ijk}={\bf S}_i \cdot({\bf S}_j \times {\bf S}_k)$ 
being nonzero only in non-planar states.
This kind of order parameter was widely discussed for the $J_1-J_2$ model
\cite{squa_ref0,squa_ref2}. We choose 
$j \in B$ as the running index and consider for 
$i,j,k$ sites forming an equilateral triangle
like sites 1, 23, 24 in Fig. \ref{fig3}, i.e. we have $k=j+{\bf a}_2$ and
$i=j+{\bf a}_1 + \frac{1}{2}{\bf a}_1$.
Then we use as order parameter (cf. \cite{squa_ref2}) 
\begin{equation} \label{chir}
W= \left (\frac{1}{N} \sum_{j \in B} \tau_j W_{ijk}\right )^2
\end{equation}
where $\tau_j$ is a staggered factor being $+1$ on sublattice $B_1$ 
(i.e. sites
17, 19, 21, 23 in Fig. \ref{fig3}) and $-1$ on sublattice $B_2$ (i.e. sites
18, 20, 22, 24 in Fig. \ref{fig3}).
As shown by Fig. \ref{figed5} this chirality is indeed large 
in the classical nonplanar states but 
we do not see significantly enhanced chiral correlations in the quantum 
ground state.

Further increasing $J_{AB}$ leads to a
transition from the complex singlet $S=0$ phase  directly to an $S=2$ phase 
at $J_{AB} \approx 3.13$, which is the quantum counterpart to the classical
canted ferrimagnetic phase IV. 
This transition is very close to the
classical transition III-IV  which is also a transition from zero
$S_{total}$ to finite $S_{total}$. 

The last transition is that to the collinear ferrimagnetic state with $S=4$ 
at $J_{AB} \approx 3.61$. This value is significantly smaller than the
corresponding classical value, again indicating that quantum fluctuations
favour collinear spin ordering leading to an enlarged stability region of the
collinear ferrimagnetic phase. 
Notice, that the additional jump just before the last transition 
is attributed to a change in total spin $S$ from $S=2$ to $S=3$
corresponding to the increase of $S_{total}$ in the classical phase IV.
One characteristics of both ferrimagnetic ($S > 0$) phases are the positive
correlations within a subsystem (see Figs. \ref{figed2} and \ref{figed3})
but negative correlations between the subsystems (see Fig. \ref{figed4}).

\section{Linear spin-wave theory for the N\'{e}el phase}
\label{IV}
The parameters for which the  N\'{e}el phase I is realized most likely
correspond to the situation in  $(Ba,Sr)_2Cu_3O_4Cl_2$. Therefore we present 
a more detailed analysis of the magnetic ordering of this phase  
using a linear spin-wave theory. Within this approach we calculate
the excitation spectrum, the order parameter as well as the spin-wave
velocity.

As usual we 
perform Holstein-Primakoff
transformation. 
Because the magnetic unit
cell contains six spins we need at least six different types of magnons 
being distinguished by a running index as illustrated in
(Fig.\ref{fig1b}). After transforming into the ${\bf {k}}$-space the
Hamiltonian (\ref{eq1}) reads
\begin{displaymath}
H=-4NJ_{AA}s^{2}-2NJ_{BB}s^{2}+\sum_{{\bf {k}}}H_{{\bf {k}}}
\end{displaymath}
with
\begin{eqnarray}
H_{{\bf {k}}}&=&4J_{AA}s\left(a^{+}_{1{\bf {k}}}a_{1{\bf {k}}}+a^{+}_{2{\bf
{k}}}a_{2{\bf {k}}}+a^{+}_{3{\bf {k}}}a_{3{\bf {k}}}+a^{+}_{4{\bf
{k}}}a_{4{\bf {k}}}\right)+4J_{BB}s\left(a^{+}_{5{\bf {k}}}a_{5{\bf {k}}}+a^{+}_{6{\bf
{k}}}a_{6{\bf {k}}}\right)\nonumber\\
&-&J_{AA}s\left(\gamma_{1{\bf {k}}}\gamma_{2{\bf {k}}}+\gamma^{\ast}_{1{\bf
{k}}}\gamma^{\ast}_{2{\bf {k}}}\right)\left(a^{+}_{1{\bf {k}}}a^{+}_{4-{\bf
{k}}}+a^{+}_{2{\bf {k}}}a^{+}_{3-{\bf
{k}}}+a_{1{\bf {k}}}a_{4-{\bf
{k}}}+a_{2{\bf {k}}}a_{3-{\bf
{k}}}\right)\nonumber\\
&-&J_{AA}s\left(\gamma_{1{\bf {k}}}\gamma^{\ast}_{2{\bf {k}}}+\gamma^{\ast}_{1{\bf
{k}}}\gamma_{2{\bf {k}}}\right)\left(a^{+}_{1{\bf {k}}}a^{+}_{3-{\bf
{k}}}+a^{+}_{2{\bf {k}}}a^{+}_{4-{\bf
{k}}}+a_{1{\bf {k}}}a_{3-{\bf
{k}}}+a_{2{\bf {k}}}a_{4-{\bf
{k}}}\right)\nonumber\\
&-&J_{BB}s\left(\gamma^{2}_{1{\bf
{k}}}+\gamma^{2}_{2{\bf {k}}}+\gamma^{\ast 2}_{1{\bf
{k}}}+\gamma^{\ast 2}_{2{\bf {k}}}\right)\left(a^{+}_{5{\bf {k}}}a^{+}_{6-{\bf
{k}}}+a_{5{\bf {k}}}a_{6-{\bf
{k}}}\right)\nonumber\\
&+&J_{AB}s\left(g+1\right)\left(\gamma_{1{\bf k}}\left(a^{+}_{1{\bf {k}}}a_{6{\bf
{k}}}+a_{2{\bf {k}}}a^{+}_{6{\bf
{k}}}\right)+\gamma^{\ast}_{1{\bf k}}\left(a_{1{\bf {k}}}a^{+}_{6{\bf
{k}}}+a^{+}_{2{\bf {k}}}a_{6{\bf
{k}}}\right)\right)/2\nonumber\\
&+&J_{AB}s\left(g-1\right)\left(\gamma_{1{\bf k}}\left(a^{+}_{1{\bf
{k}}}a^{+}_{6{-\bf
{k}}}+a_{2{\bf {k}}}a_{6-{\bf
{k}}}\right)+\gamma^{\ast}_{1{\bf k}}\left(a_{1{\bf {k}}}a_{6-{\bf
{k}}}+a^{+}_{2{\bf {k}}}a^{+}_{6-{\bf
{k}}}\right)\right)/2\nonumber\\
&+&J_{AB}s\left(g+1\right)\left(\gamma_{2{\bf k}}\left(a^{+}_{4{\bf
{k}}}a_{5{\bf
{k}}}+a_{3{\bf {k}}}a^{+}_{5{\bf
{k}}}\right)+\gamma^{\ast}_{2{\bf k}}\left(a_{4{\bf {k}}}a^{+}_{5{\bf
{k}}}+a^{+}_{3{\bf {k}}}a_{5{\bf
{k}}}\right)\right)/2\nonumber\\
&+&J_{AB}s\left(g-1\right)\left(\gamma_{2{\bf k}}\left(a^{+}_{4{\bf
{k}}}a^{+}_{5{-\bf
{k}}}+a_{3{\bf {k}}}a_{5-{\bf
{k}}}\right)+\gamma^{\ast}_{2{\bf k}}\left(a_{4{\bf {k}}}a_{5-{\bf
{k}}}+a^{+}_{3{\bf {k}}}a^{+}_{5-{\bf
{k}}}\right)\right)/2\nonumber\\
&-&J_{AB}s\left(g-1\right)\left(\gamma_{1{\bf k}}\left(a^{+}_{2{\bf
{k}}}a_{5{\bf
{k}}}+a_{1{\bf {k}}}a^{+}_{5{\bf
{k}}}\right)+\gamma^{\ast}_{1{\bf k}}\left(a_{2{\bf {k}}}a^{+}_{5{\bf
{k}}}+a^{+}_{1{\bf {k}}}a_{5{\bf
{k}}}\right)\right)/2\nonumber\\
&-&J_{AB}s\left(g+1\right)\left(\gamma_{1{\bf k}}\left(a^{+}_{2{\bf
{k}}}a^{+}_{5{-\bf
{k}}}+a_{1{\bf {k}}}a_{5-{\bf
{k}}}\right)+\gamma^{\ast}_{1{\bf k}}\left(a_{2{\bf {k}}}a_{5-{\bf
{k}}}+a^{+}_{1{\bf {k}}}a^{+}_{5-{\bf
{k}}}\right)\right)/2\nonumber\\
&-&J_{AB}s\left(g-1\right)\left(\gamma_{2{\bf k}}\left(a^{+}_{3{\bf
{k}}}a_{6{\bf
{k}}}+a_{4{\bf {k}}}a^{+}_{6{\bf
{k}}}\right)+\gamma^{\ast}_{2{\bf k}}\left(a_{3{\bf {k}}}a^{+}_{6{\bf
{k}}}+a^{+}_{4{\bf {k}}}a_{6{\bf
{k}}}\right)\right)/2\nonumber\\
&-&J_{AB}s\left(g+1\right)\left(\gamma_{2{\bf k}}\left(a^{+}_{3{\bf
{k}}}a^{+}_{6{-\bf
{k}}}+a_{4{\bf {k}}}a_{6-{\bf
{k}}}\right)+\gamma^{\ast}_{2{\bf k}}\left(a_{3{\bf {k}}}a_{6-{\bf
{k}}}+a^{+}_{4{\bf {k}}}a^{+}_{6-{\bf
{k}}}\right)\right)/2\nonumber\\
\label{eq6}
\end{eqnarray}
and $\gamma_{n{\bf {k}}}=\exp\left(i{\bf {k}}{\bf {a}}_{n}/2\right)$. Here
$N$ is the number of geometrical unit cells $N={\cal N}/3$.
The
vectors ${\bf {a}}_{n}$ are the unit vectors of the geometrical
lattice: ${\bf {a}}_{1}=a\left(1,0\right)$ and ${\bf
{a}}_{2}=a\left(0,1\right)$ and $a$ is the lattice constant (see Fig.
\ref{fig1}a). $g$ is defined as
$g=\cos\left(\varphi\right)$, where $\varphi$  parametrizes
the angle between the N\'{e}el
states of the  classical subsystems A and B. 
Without any further calculation it is obvious that quantum
fluctuations stabilize collinear ordering. According to the Hellmann-Feynman 
theorem 
\cite{ref11} the relation $\partial E/\partial \lambda=\left\langle
\partial H/\partial \lambda\right\rangle$ holds, 
where $H$ is a Hamiltonian depending on
a parameter $\lambda$ and $E$ is an eigenvalue of $H$. 
Because (\ref{eq6}) depends
on $\cos\left(\varphi\right)$-terms only one finds $\partial E/\partial
\varphi\sim\sin\left(\varphi\right)$ being zero for $\varphi=0,\pi$, i.e. as
discussed already above in
the quantum system the classical degeneracy is lifted and collinear spin
structures are preferred.
Thus, all quantities have to be calculated as averages over both possible
ground states belonging to $\varphi=0$ and $\pi$. 

The diagonalization of the bosonic Hamiltonian is carried out as usual by
means of Green
functions. As it should be there are six non-degenerated spin-wave branches - two of them
are optical 
whereas the remaining ones 
are two acoustical branches per subsystem. The acoustical
branches become zero in the center of the Brillouin zone, only.  
Expanding these branches in the vicinity of ${\bf {k}}=0$
gives two different spin-wave velocities $c_{A}$ and $c_{B}$. 
\begin{eqnarray}
c_{A}&=&as\sqrt{2J_{AA}^{2}+4J_{BB}^{2}-\frac
{J_{BB}J_{AB}^{2}}{J_{AA}}+q}\; \; , \; \;
c_{B}=as\sqrt{2J_{AA}^{2}+4J_{BB}^{2}-\frac {J_{BB}J_{AB}^{2}}{J_{AA}}-q} \;\; ,
\nonumber\\
q&=&\sqrt{\left(2J_{AA}^{2}-4J_{BB}^{2}\right)^{2}+\frac
{8J_{AB}^{2}}{J_{AA}^{2}}\left(J_{AA}^{3}J_{BB}-J_{AA}J_{BB}^{3}\right)+\frac
{J_{AB}^{4}}{J_{AA}^{2}}\left(J_{BB}^{2}-J_{AA}^{2}\right)} \; \;,
\label{eq7}
\end{eqnarray}
where the two acoustical branches
belonging to the same subsystem have identical spin-wave velocities.
At the classical phase-transition point  $J_{AB}=2\sqrt{J_{AA}J_{BB}}\;\;$ 
$c_{B}$ becomes  zero wheras $c_A$ remains
finite.\\
The ground-state energy $E_{0}^{{\em {sw}}}$ is given by
\begin{equation}
E_{0}^{{\em {sw}}}=-2N\left(2J_{AA}+J_{BB}\right)s\left(s+1\right)+\sum_{{\bf
{k}}}\sum_{m=1}^{6}\omega_{m{\bf {k}}}/2
\label{eq8}
\end{equation}
and the sublattice magnetizations is calculated by
\begin{equation}
\left\langle S_{n}^{z} \right\rangle=s-\frac {2}{N}\sum_{{\bf
{k}}}\left\langle a^{+}_{n{\bf {k}}}a_{n{\bf {k}}}\right\rangle \; ,\quad
n=1,\ldots ,6
\label{eq9}
\end{equation}  
for $A$-spins as well as for $B$-spins. 

\begin{figure}
\begin{center}
\epsfig{file=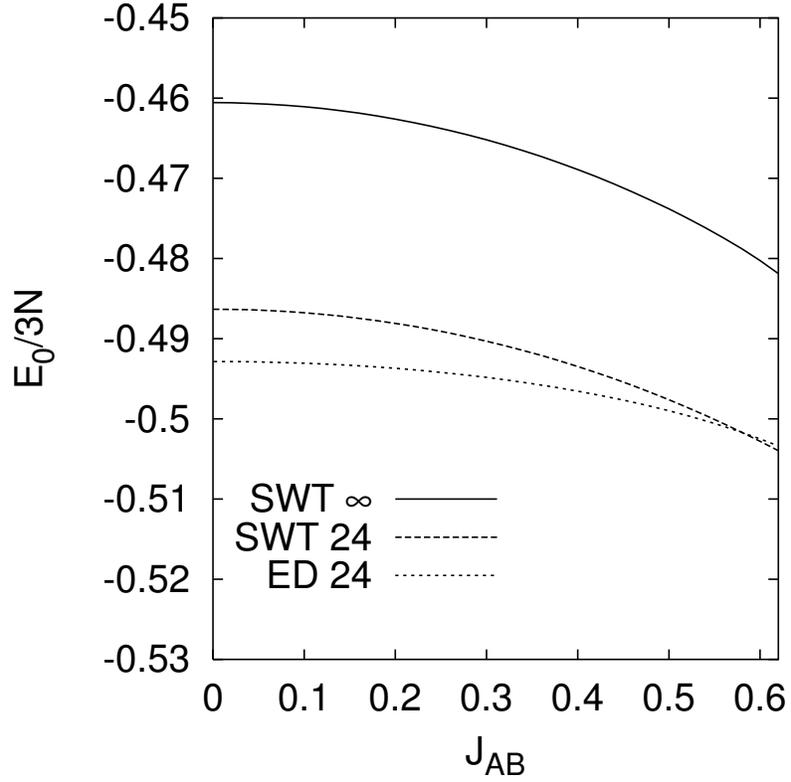,scale=0.6,angle=270.0}
\end{center}
\caption{Ground-state energy per spin for $s=1/2$: 
spin-wave (${\cal N}=\infty$ and ${\cal N}=24$) and
exact-diagonalization results and (${\cal N}=24$).\label{figsw1}}
\end{figure}

\begin{figure}
\begin{center}
\epsfig{file=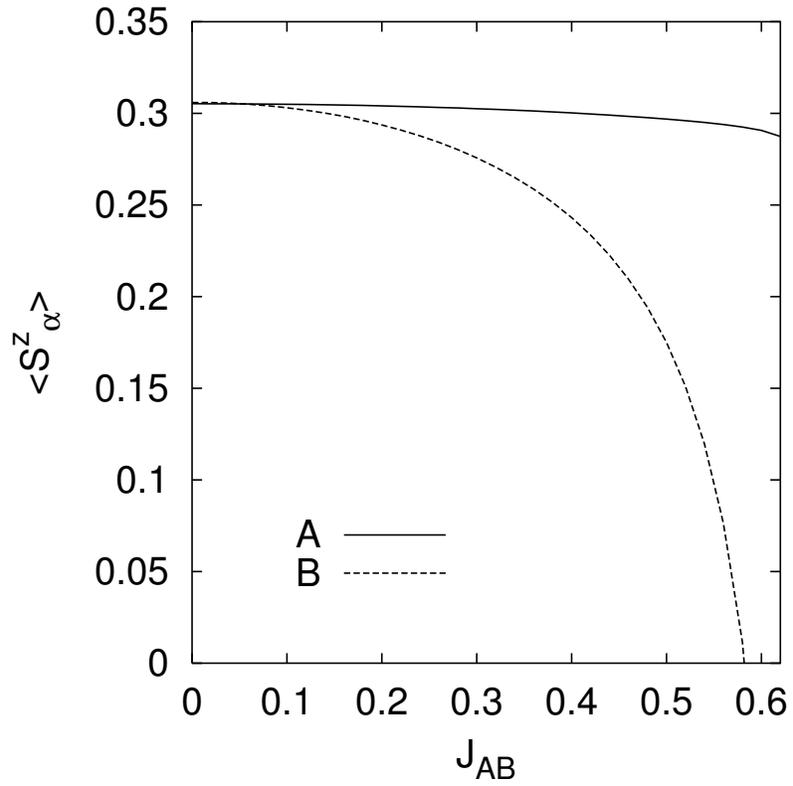,scale=0.6,angle=270.0}
\end{center}
\caption{Sublattice magnetizations $\langle S_A^z \rangle $ and 
$\langle S_B^z \rangle $ (see eq. (\ref{eq9}))
for $s=1/2$ in the infinite system: 
spin-wave results ($J_{AA}=1$, $J_{BB}=0.1$).\label{figsw2}}
\end{figure}

The results of the spin-wave calculation for the relevant parameters
$J_{AA}=1$ and $J_{BB}=0.1$ are presented in Figs. \ref{figsw1} and 
\ref{figsw2}.
We start with the ground-state energy shown in Fig.\ref{figsw1}. While the classical
energy in phase I is independent of $J_{AB}$ we find a slight decrease with
$J_{AB}$ in the quantum model. For comparison we show the
exact-diagonalization and the spin-wave results for
${\cal N}=24$. The difference is small ($1.3 \%$ for $J_{AB}=0$) 
indicating that
linear spin-wave theory seems to be well appropriate for phase I.

\begin{figure}
 \begin{center}
 \epsfig{file=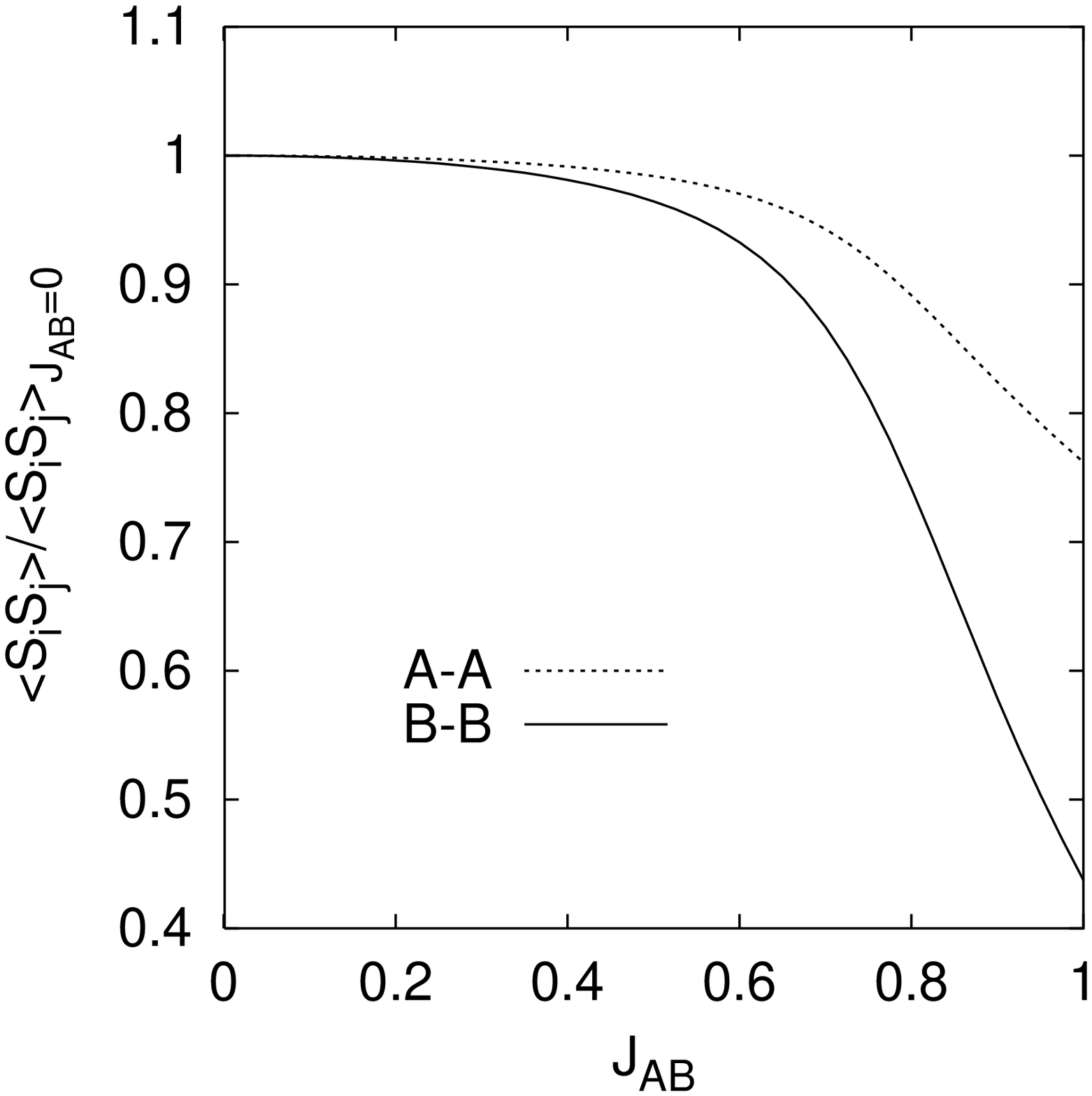,scale=0.6,angle=270.0}
 \end{center}
 \caption{Nearest-neighbour correlation within the A subsystem and the B
 subsystem (exact diagonalization results for ${\cal N}=24$, 
 $J_{AA}=1$, $J_{BB}=0.1$). 
 For better comparison we have scaled $\langle {\bf S}_i {\bf S}_j \rangle$
 by its
 corresponding values for $J_{AB}=0$.
 \label{figed6}}
\end{figure}

The spin-wave theory allows to calculate the corresponding 
sublattice magnetizations $\langle S^z_A \rangle$ and $\langle S^z_B \rangle$ 
in the $A$ and $B$ subsystems for $\cal N \to \infty$
(see eq. \ref{eq9}). The results are shown in Fig. \ref{figsw2}. 
Although $\langle S^z_A \rangle$ is slightly diminished with growing
$J_{AB}$ the N\'eel order of
the $A$ subsystem is stable within the limits of the classical phase I.
Contrary to that the N\'{e}el
order 
of the $B$ subsystem is stronger suppressed and breaks down at $J_{AB} \approx
0.58$. 
This finding is supported by the ED results for the spin-spin
correlations (see Fig. \ref{figed6}),
where we see also a
stronger suppressing of B-B correlations with growing $J_{AB}$ than of A-A
correlations. Hence we argue that the strong quantum fluctuations in the
spin-half model in combination with strong frustration may lead to 
a novel
ground-state phase with N\'{e}el ordering in the $A$ subsystem but quantum
disorder in the B subsystem. A similar observation recently has been made
for the frustrated square-lattice $J_1-J_2$ spin-one spin-half 
ferrimagnet, where for strong frustration the spin-half subsystem might be
disordered but the spin-one subsystem is ordered \cite{ferri}.

\section{Summary} 
\label{V}
In this paper the results of exact
diagonalization and linear spin-wave theory  for the ground state of a 
system of two interpenetrating spin-half
Heisenberg antiferromagnets on square lattices are presented. We consider
intra-subsystem couplings of different strength $J_{BB} = 0.1J_{AA}$ which
correponds to the situation in $(Ba,Sr)_2Cu_3O_4Cl_2$. 
In addition to
strong quantum fluctuations there is a competing inter-subsystem 
coupling $J_{AB}$ between
both spin systems giving rise to interesting frustration effects.
The classical version of our model possesses a rich magnetic phase diagram
with collinear, planar and non-planar ground states. 
Quantum fluctuations may change the ground-state phases. In particular, we
find indications for preferring collinear versus non-collinear 
and planar versus non-planar phases by 
quantum fluctuations. 

For small $J_{AB}$ both spin subsystems are in the N\'{e}el
state. These N\'{e}el
states decouple classically. Quantum fluctuations lead to a fluctuational
coupling of both subsystems.
With increasing $J_{AB}$ the frustration tends to destroy the N\'{e}el
ordering of the weaker coupled $B$ subsystem but not 
in the stronger
coupled $A$ subsystem. The comparison between exact finite-size
data and approximate
spin-wave data gives a good agreement between both approaches.   \\

{\bf Acknowledgments}\\
This work was  supported by 
the Deutsche Forschungsgemeinschaft (Grant No. Ri615/7-1).

\begin{table}
\begin{center}
\begin{tabular}{|c|c|c|c|c|c|}
&&&&&\\
phase & range of stability & energy $\frac{E}{s^2{ \cal N}}$ & $\frac{S_A}{s2N}$ & $\frac  {S_B}{sN}$ &
$\frac{S_{total}}{s{\cal N}}$\\
&&&&&\\
\hline \hline
&&&&&\\
I & $0\leq J_{AB}\leq 2\sqrt{J_{AA}J_{BB}}$ & $-\frac{2}{3}J_{BB}-\frac{4}{3}J_{AA}$ & 0 & 0 & 0 \\
&&&&&\\
II & $2\sqrt{J_{AA}J_{BB}}\leq J_{AB}\leq 2\sqrt{2}J_{AA}$ & $-\frac{1}{6}\frac{J_{AB}^2}{J_{AA}}-\frac{4}{3}J_{AA}$ & 0 & 0 & 0\\
&&&&&\\
III & $2\sqrt{2}J_{AA}\leq J_{AB}\leq (\ref{IV_7})$ & $-\frac{1}{\sqrt{2}}\frac{4}{3}J_{AB}$ & 0 & 0 & 0\\
&&&&&\\
IV & $(\ref{IV_7})\leq J_{AB}\leq (\ref{IV_8})$ & $E_{IV}$ from eq. (\ref{IV_1}) &
eq. (\ref{IV_2}) & eq. (\ref{IV_2}) & eq. (\ref{IV_2})\\
&&&&&\\
V & $(\ref{IV_8})\leq J_{AB}$ & $-\frac{4}{3}J_{AB}+\frac{2}{3}J_{BB}+\frac{4}{3}J_{AA}$ & 1 & 1 & 1/3\\
&&&&&\\
\end{tabular}
\end{center}
\caption{\label{tab1} The five classical ground-state phases for $J_{AB},
J_{BB}>0$ and $J_{BB}\le J_{AA}$, where $S_{A(B)}=|\sum_{n\in A(B)} {\bf S}_i|$ 
is the total spin of 
subsystem
A(B), and $S_{total}=|\sum_{n} {\bf S}_i|$ is the total spin of the whole system.}
\end{table}

%


\begin{thebibliography}{99}
\bibitem{ref4} S.Noro, H.Suzuki, T. Yamadaya, Solid State Commun. {\bf 76},
711 (1990);
S.Noro et al., Mater.Sci.Eng. B {\bf 25} (1994), 167.
\bibitem{ref5} K. Yamada, N. Suzuki, J. Akimitsu, Physica B {\bf 213-214},
191 (1995).
\bibitem{ref3} F.C. Chou {\em et al.}, Phys. Rev. Lett. {\bf 78}, 535
(1997).
\bibitem{ref1} Y.J. Kim {\em et al.}, Phys. Rev. B {\bf 64}, 024435 (2001).
\bibitem{ref2} A.B. Harris {\em et al.}, Phys. Rev B {\bf 64}, 024436
(2001).
\bibitem{valya} T.V.Valyanskaya and V.I.Sokolov, Zh. Eksp. Teor. Fiz. {\bf
75}, 325 (1978)
(Sov. Phys. JETP {\bf 48}, 161 (1978)).
\bibitem{shender} E.F. Shender, Zh. Eksp. Teor. Fiz. {\bf 83}, 326 (1982)
(Sov. Phys. JETP {\bf 56}, 178 (1982)).
\bibitem{brueck} Th.Br\"uckel, C.Paulsen, K.Hinrichs and W.Prandl,
Z.Phys. B {\bf 97}, 391 (1995).
\bibitem{rosner} H.Rosner, R.Hayn and J.Schulenburg, Phys. Rev B {\bf 57},
13660 (1998).
\bibitem{ref10} J. Richter, A.Voigt, J.Schulenburg, N.B. Ivanov and R.Hayn,
 J. Magn. Magn. Mater. {\bf  177-181}, 737 (1998).
\bibitem{ferri}
N.B.Ivanov, J.Richter and D.J.J.Farnell,
    Phys. Rev B {\bf 66}, 014421 (2002). 
\bibitem{villain} J. Villain, R. Bidaux, J.P. Carton, R. Conte,
J. Phys. {\bf 41}, 1263 (1980).
\bibitem{ccm} 
S.Kr\"uger and J.Richter,
       Phys. Rev. B {\bf 64}, 024433 (2001). 
\bibitem{squa_ref0} E.Dagotto and A.Moreo,  {Phys. Rev. Lett.}
  {\bf 63}, 2148 (1989).
\bibitem{squa_ref1} H.J.~Schulz and T.A.L.~Ziman, {Europhys. Lett.}
  {\bf 18}, 355 (1992).
\bibitem{squa_ref2} 
   J.~Richter, Phys. Rev. B $\bf47$, 5794 (1993).
\bibitem{squa_ref3} L.~Capriotti and S.~Sorella, 
{Phys. Rev. Lett.} {\bf 84}, 3173 (2000).
\bibitem{squa_ref4}
    O.P.~Sushkov, J.~Oitmaa and Zheng Weihong,
   {Phys. Rev. B}
   {\bf  63}, 104420 (2001).
\bibitem{ref11} R.P. Feynman, Phys. Rev. {\bf 56}, 340 (1939). 
\end{thebibliography}
\end{document}